\documentclass[aps,prl,a4paper,twocolumn]{revtex4-1}
\usepackage{bm,mathrsfs}
\usepackage{hyperref,amsmath,amssymb,bm,amsthm}
\usepackage{color,epsf,graphicx,amssymb,longtable,afterpage,mathrsfs}
\usepackage{bm,mathrsfs}
\usepackage{hyperref,amsmath,amssymb,bm,amsthm}
\usepackage{color,epsf,graphicx,amssymb,longtable,afterpage,mathrsfs}

\def\be{\begin{equation}} 
\def\ee{\end{equation}}
\def\bea{\begin{eqnarray}} 
\def\eea{\end{eqnarray}}
\def \line{\hbox to \hsize}    
\def\frac #1#2{{#1\over #2}}

\def \ket #1{{\vert #1\rangle}}
\def \bra #1{{\langle #1\vert}}

\def\1{\mbox{\bf 1}}

\newcommand{\para}[1]{}

\theoremstyle{definition} 
\newcommand{\comment}[1]{}
\begin{document}
\title{Topological pumps and adiabatic cycles}

\author{Rahul Roy}
\affiliation{Rudolf Peierls Centre for Theoretical Physics,
Oxford University \\
1 Keble Road, Oxford, OX1 3NP, UK}
\begin{abstract}

 Topological insulators have gapless states at their boundaries
while trivial insulators generically do not. We consider loops
in the spaces of Hamiltonians of topologically trivial Bloch
insulators, and show that there exist loops for which the
boundary gap must necessarily close at some point or points along
the loop. We show that some of these loops may be regarded,
depending on the symmetry class of the insulator and its
physical dimension, as defining pumps of charge, fermion
parity, and in more exotic cases of other quantities such as a
$Z_2$ parity.

\end{abstract}
\maketitle
It has recently been recognized that ordinary Bloch
     insulators with time-reversal symmetry (TRS) come in more
     than one
     variety~\cite{kane2005qsh,moore06a,PhysRevB.79.195322,fuL06a}. Since
     the first predictions and the subsequent experimental
     confirmations, the field of topological insulators (Bloch
     insulators of the non-trivial variety) has seen an
     explosion of both theoretical and experimental
     interest~\cite{2010arXiv1002.3895H}.  Topological
     insulators belong to a different phase of Bloch insulators
     than trivial insulators and it is impossible to go
     continuously from the Hamiltonian of a trivial insulator
     to that of a topological insulator without going through a
     point where the bulk gap closes. Thus, the existence of
     topological insulators for a particular symmetry class and
     physical dimension reveals an interesting fact about the
     topological structure of the corresponding space of Bloch
     Hamiltonians, namely that there are paths in this space
     along which the bulk gap must necessarily close at some
     point or points.

                  In this paper, we consider a related issue
     which is the following: Are there loops in the space of
     Bloch Hamiltonians of trivial insulators (which
     generically do not have gapless boundary states) along
     which the boundary gap (in a system with a boundary)
     necessarily closes at some point or points despite the
     bulk gap staying open throughout the loop?  A loop in the
     space of Bloch Hamiltonians for which the bulk excitation
     gap remains open may be regarded as an adiabatic cycle of
     a Bloch insulator, a term we use to describe the process
     of transforming gradually the Hamiltonian of a Bloch
     insulator by changing, for instance, the periodic
     potential.Physically,
     one can imagine an adiabatic cycle arising from a
     continuous and periodic physical perturbation such as a
     strain that leads to a distortion of the periodic
     potential and hence the band structure.We call adiabatic cycles for which
     the boundary gap must necessarily close non-trivial
     adiabatic cycles. Later, we will argue that there is
     indeed a non-trivial topological structure to these loops
     and that they may be assigned a non-trivial ``bulk''
     topological invariant which may be determined from the
     bulk wavefunctions when periodic boundary conditions are
     used.

             We will see that, though not originally studied
     from the perspective of a closing of the boundary gap in a
     finite system, the charge pumps in one dimension (1d)
     introduced by Thouless~\cite{Thouless_83_6083} are
     examples of non-trivial adiabatic cycles in the sense
     discussed above, i.e., they are accompanied by a closing
     of the boundary gap in a semi-infinite system with an
     boundary.  In the case of the Thouless charge pump, we
     shall see that the total fermion number or charge of the
     ground state in a semi-infinite system changes along with
     a closing of the gap.  For a 1d system with two
     boundaries, this leads to the picture of charge entering
     through one boundary and leaving through the other in the
     course of a cycle. Thus, the adiabatic cycle may be
     regarded as a pumping of charge.  A similar process occurs
     for some other classes of non-trivial adiabatic cycles
     that we study. In some of these cases, quantum numbers
     such as a Hall conductance, or spin Hall conductance may
     be associated with the states that cross the Fermi
     energy. Thus the corresponding adiabatic cycles may be
     regarded as ``Hall pumps'' or ``spin Hall'' pumps.

                 The remainder of the paper is organized as
     follows. We first provide a topological classification of
     adiabatic cycles based on their bulk Bloch wavefunctions using
     known results of K-theory. We then go on to argue that
     adiabatic cycles with a non-trivial bulk topological invariant
     lead to a closing of the gap in systems with an
     boundary. Thereafter, we discuss a number of interesting examples
     of topological adiabatic cycles and their
     boundary invariants and pumping properties.

     We begin with a topological classification of adiabatic
     cycles.  Our primary interest is in the case of insulators
     belonging to the trivial class. Since, however, it is no
     harder to do so, we will also consider cases where the
     insulators are topologically non-trivial.  The
     classification of adiabatic cycles is based on questions
     similar to those that led to a classification of
     insulators. Are there adiabatic cycles which cannot be
     deformed into each other by a continuous deformation of
     the band structure? It is clear that quantities such as
     the number of occupied bands of a band insulator and its
     topological class are invariants of an adiabatic cycle. A
     complete classification of all such invariants can be obtained
     using the tools of K-theory which have recently led to a
     fuller understanding of the periodic structure in the
     classification of topological
     insulators~\cite{Kitaev_0901.2686}. Our focus is primarily
     on a particular class of invariants of the adiabatic
     cycles, which we call the strong invariants. There is a
     chain in dimensions in the structure of the topological
     invariants. Invariants associated with adiabatic cycles of
     d-dimensional insulators always lead to invariants of
     adiabatic cycles of d+1-dimensional insulators. This
     situation is reminiscent of topological insulators in 3d,
     where there are 4 invariants, three of which may be
     thought of as 2d or ``weak'' invariants, and one of them
     is an intrinsically 3d or ``strong'' invariant. In the
     current situation, both the topological invariant of the
     insulator as well as the lower dimensional invariants of
     the adiabatic cycles can be regarded as weak invariants.

   We now explicitly identify a group associated with the
     strong invariants of adiabatic cycles of Bloch insulators
     without TRS. In the tight-binding representation, the
     Hamiltonian for electrons in a periodic potential can be
     represented by a Hermitian matrix at each point in the
     Brillouin zone. Setting the chemical potential to zero,
     this matrix has non-vanishing determinant in a Bloch
     insulator. The Brillouin zone for a d-dimensional
     insulator is topologically equivalent to the d-torus,
     $T^d$ and a loop, to the unit circle, $S^1$.  Thus, an
     adiabatic cycle defines a smooth map from $T^{d}\times
     S^{1}$ to the set, $\mathscr{H}(d,1)$, of Hermitian
     matrices which have non-zero determinant.  An equivalence
     relation can be defined for these maps which isolates the
     strong topological invariant of the adiabatic cycles in
     much the same way as in the case of topological
     insulators~\cite{Kitaev_0901.2686}. Further, we can define
     an operation which, given two maps, $H_1, H_2$ from
     $T^{d}\times S^{1}$ to $\mathscr{H}(d,1)$, produces a
     third map whose value at a given point $(k,t)$ in
     $T^{d}\times S^{1}$ is the direct sum, $H_1(k,t)\oplus
     H_2(k,t)$, of the matrices, $H_1(k,t), H_2(k,t)$ obtained
     from the two maps, $H_1$ and $H_2$. The set of the
     equivalence classes of maps defined through the
     equivalence relation mentioned earlier forms a group, $G$,
     under this operation, which is in fact isomorphic to a
     group denoted by $\tilde{K}(S^{d+1})$ in K-theory. It is a
     standard result in K-theory that this group is the set of
     integers for odd $d$ and the trivial group for even
     $d$~\cite{Karoubi}. 

     The symmetries of the insulator play a crucial part in the
     classification of their adiabatic cycles by imposing
     constraints on the corresponding maps from $T^{d}\times
     S^1 $ to $\mathscr{H}(d,1)$. Each fundamental set of
     symmetries (such as TRS and particle-hole symmetry) of an
     insulator corresponds to one of the ten Altland Zirnbauer
     (AZ) symmetry classes of random
     matrices~\cite{Altland_97_1142}.  Using the constraints
     imposed by symmetries and the procedure outlined above,
     one can identify the appropriate $K$ groups for the
     classification of the strong invariants of adiabatic
     cycles in all the AZ symmetry classes with the help of
     standard results in
     K-theory~\cite{Karoubi,roy_unpublished}. This is analogous
     to the classification of topological insulators and
     defects using K-theory which has been outlined in several
     recent
     works~\cite{Kitaev_0901.2686,Stone_1005.3213,Teo_1006.0690}.
     The results of the classification are presented in
     Table~\ref{tbl:results}.

           \begin{table}[htb]
           \caption{\label{tbl:results} Classification of adiabatic cycles
           of Bloch insulators. Each column corresponds to a particular
           symmetry class identified by its Cartan label with the
           different rows corresponding to different physical dimensions.}
           \begin{center}
           \begin{tabular}{|c|c|c|c|c|c|c|c|c|c|c|}
             \hline
             d & A & AIII & AI & BDI & D & DIII & AII & CII & C & CI \\
             \hline
             0 & 0 & $Z$ & $Z_2$ & $Z_2$ & 0 & $Z$ & 0 & 0 & 0 & $Z$ \\
             1 & $Z$ & 0 & $Z$ & $Z_2$ & $Z_2$ & 0 & $Z$ & 0 & 0 & 0 \\
             2 & 0 & $Z$ & 0 & $Z$ & $Z_2$ & $Z_2$ & 0 & $Z$ & 0 & 0 \\
             3 & $Z$ & 0 & 0 & 0 & $Z$ & $Z_2$ & $Z_2$ & 0 & $Z$ & 0 \\
             \hline
           \end{tabular}
         \end{center}
       \end{table}

     We now relate the existence of distinct equivalence
     classes of adiabatic cycles to the closing of the boundary gap.
     Consider an infinite $d$-dimensional ($d\ge 1$) trivial
     insulator which we regard as two semi-infinite insulators,
     labeled left and right, joined at the $d-1$ dimensional
     plane, $x=0$.  Let us now imagine that we smoothly change
     the Hamiltonian of the left insulator as a function of a
     parameter, $t$, along the path of a non-trivial topological
     adiabatic cycle while the Hamiltonian of the right is kept
     constant. At $t=0$, the Hamiltonian is the same for the left
     and the right insulators. For small $t$, one can expect to
     vary the Hamiltonian in the vicinity of $x = 0$ without
     forming localized sub-gap states in the vicinity of this
     plane.  However, it is clear that this can not be done for
     all $t$. Indeed, if it could, it would define a continuous
     deformation from the adiabatic cycle on the left to the
     constant or trivial loop on the right. Since the cycle on
     the right is a topological adiabatic cycle, and, by
     definition, one which cannot be continuously deformed to a
     constant one, this provides a contradiction, thus proving
     the existence of gapless boundary states. These
     boundary modes are expected to persist when one replaces
     the trivial insulator with the vacuum.

     \textbf{d=1:}  Let us now study the simplest examples of
     non-trivial adiabatic cycles which are accompanied by a
     closing of the boundary gap, namely the charge pumps of 1d
     insulators without TRS, which were studied by Thouless in
     the context of quantized fermion number
     transport~\cite{Thouless_83_6083}. Consider a
     semi-infinite 1d insulator which extends along the
     positive x-axis, whose Hamiltonian, $H_{1d}(t)$, is varied
     along the path of a topological adiabatic cycle such that
     $H_{1d}(2\pi) = H_{1d}(0)$. This cycle can be mapped onto
     the Hamiltonian, $H_{2d}$, of a 2d Bloch insulator on a
     cylinder, by writing $ H_{2d} = \sum_{k_y} H_{1d}(t=k_y)$,
     where $k_y$ is the scaled momentum along the circumference
     of the cylinder. From the defining property of a 1d
     charge pump, the 2d Bloch insulator displays the integer
     quantum Hall effect. Consequently, it must have
     boundary modes on account of Laughlin's gauge
     argument~\cite{Laughlin_81_5632} and so must the adiabatic
     cycle. Let $\{t_i\}$ be the set of points at which gapless
     zero energy states occur. In the vicinity of these points,
     one may define boundary Hamiltonians, $
     H_{\textrm{b}}(t) = \sum_{i} E_i
     \ket{u_i(t}\bra{u_i(t)} $, where the sum is over all
     states that decay exponentially away from the
     boundary. One can always ensure (through slight
     deformations of the Hamiltonian, if necessary) that there
     exists some interval, $[t_i - \epsilon, t_i + \epsilon]$,
     around each of the $t_i$ such that no extended states join
     the set of localized boundary states in this interval and
     that the boundary Hamiltonian has a gap (albeit small) at
     the end points of the interval.

     Let $n_i ^{\pm}$ be the charges associated with the ground
     states of the boundary Hamiltonians, $H_{\textrm{b}}(t_i
     \pm \epsilon)$. Then, the net spectral flow associated
     with the cycle is equal to $\sum_i n_i^{+} - \sum_i n_i
     ^{-}$. It cannot change by arbitrary deformations of the
     boundary spectrum and is hence an invariant of the
     cycle. Using Laughlin's gauge argument (which proves the
     correspondence between the Chern invariant and spectral
     flow for the 2d Bloch insulator) and the correspondence
     between the 1d and the 2d systems, it is easy to prove
     that this boundary invariant must equal the bulk
     topological invariant of the adiabatic cycle. The flow at
     the gapless points in the boundary spectrum leads to a
     flow of fermions from the occupied states to the
     unoccupied states in a system with a fixed number of
     particles. (In a system with a fixed chemical potential,
     it leads to a flow of particles in and out of the system.)
     Thus the adiabatic cycle can be regarded as a fermionic
     number pump, or a charge pump for a system with charged
     fermions.

                Adiabatic cycles of 1d insulators in classes AI
     and AII are also classified by an integer (Z)
     invariant. In both of these cases, the first Chern number
     provides the topological invariant, and there is a net
     spectral flow and consequently a pump of fermion
     number. Time-reversal symmetry ensures that the net
     fermion number pumped in the case of Class AII is always
     an even integer.

        Further interesting cases are the $Z_2$ adiabatic
     cycles of insulators in classes BDI and D. Consider a
     topological adiabatic cycle of a semi-infinite 1d trivial
     superconductor. If at a degeneracy point, $t_0$, a single
     pair of boundary states crosses zero, the parity of the
     ground state changes~\footnote{By a suitable deformation
     of the cycle, one may ensure that a simultaneous
     degeneracy at the other end of the system does not
     occur.}.  The net change in parity across all the gapless
     points is an invariant of the cycle and is non-trivial for
     a cycle with a non-trivial $Z_2$ invariant, $E$. The fermion
     parity pump studied above is qualitatively different from
     the Josephson junction of a 1d topological superconductor
     studied before~\cite{Kitaev_01_131,Teo_1006.0690} where no
     spectral flow occurs when open boundary conditions are
     used.

      \textbf{d=2:}  We next consider insulators in two spatial
     dimensions and argue that the non-trivial cycles in
     classes D and DIII and those characterized by an odd
     integer in class BDI can be regarded as parity pumps.  Let
     us first examine adiabatic cycles of insulators in Class
     D. Consider a 2d crystal with lattice basis vectors
     $(\bm{a}=a\hat{y},\bm{b})$ which has periodic boundary
     conditions and an even number of unit cells in the
     y-direction (for convenience of discussion) and is
     semi-infinite along $\bm{b}$.  At each point of time, t,
     the Hamiltonian can be written as a sum: $ {\cal H}(t) = H(0,t) +
     H(\pi,t) + \sum_{0<k_y<\pi}( H(k_y,t)+H(-k_y,t)) $ where
     $0,\pi,k_y$ are the crystal momenta in the $y$-direction,
     rescaled so that they lie between $-\pi$ and $\pi$.  Under
     a particle-hole transformation, a general crystal momentum
     $k_y$ goes to $-k_y$, while the momenta $0$ and $\pi$ are
     invariant. Thus, the operators $H(0,t), H(\pi,t)$ and
     $H'(k_y,t)=H(k_y,t)+H(-k_y,t)$ for $0 <k_y<\pi$ can all be
     regarded as defining adiabatic cycles of 1d
     superconducting Hamiltonians in class D and one may
     therefore associate $Z_2$ invariants, $E(0),E(\pi)$, and
     $E(k_{y})$ with them. It is clear that $E(k_{y})$ for
     $0<k_{y}< \pi$ is always the trivial element of the group
     $Z_2$. The topological invariant of an adiabatic cycle of
     a 2d superconductor is the product of the topological
     invariants $E(0)E(\pi)$. Since the net change in parity
     for the Hamiltonians, $H'(k_{y})$, is always trivial, this
     product also determines the net change through spectral
     flow of the parity of a lattice superconductor with an
     even number of unit cells around the cylinder. A more
     general statement which is independent of the width of the
     cylinder is that the product of the net changes in parity
     for an adiabatic cycle in the presence and absence of a
     $\pi$ flux along the axis of the cylinder is equal to the
     $Z_2$ invariant.  Thus the non-trivial adiabatic cycles
     for 2d insulators in class D can be interpreted as parity
     pumps. From similar considerations, it is easy to show
     that the adiabatic cycles in class BDI with a topological
     invariant $\nu$ such that $|\nu|$ is an odd integer also
     correspond to parity pumps.

                Non-trivial adiabatic cycles in class DIII in
     2d can be regarded as spin or $Z_{2}$ parity pumps. For
     these pumps, one can write the ground state at any given
     moment as a product
     $\ket{\Omega_{I}(t)}\ket{\Omega_{II}(t)}$ where
     $\ket{\Omega_{I}(t)}$ gets mapped onto
     $\ket{\Omega_{II}(t)}$ under time-reversal symmetry.  One
     can then define the associated parities, $P_{I}(t),
     P_{II}(t)$, such that the total parity is $P(t) =
     P_{I}(t)P_{II}(t)$ and while the net change in the parity
     of the entire system during the cycle is zero, there is a
     net change in the parities, $P_{I}(t)$ and $P_{II}(t)$ for
     non-trivial adiabatic cycles of insulators with an even
     number of unit cells in the periodic direction. 
                    
     \textbf{d=3:} Similar arguments to those used above can
     be used to prove that adiabatic cycles of
     superconductors in 3d in class D with a ``strong'' 3d
     invariant $\nu \in Z$ such that $\nu$ is odd can be
     regarded as parity pumps. Similarly, $Z_2$ adiabatic
     cycles of superconductors in class DIII with a
     non-trivial strong $Z_2$ invariant can be regarded as
     $Z_2$ parity pumps. In these cases, the net parity
     change occurs in systems with periodic boundary
     conditions and an even number of unit cells along two of
     the lattice basis vectors $\bm{a},\bm{b}$ and open
     boundary conditions along a third basis vector,
     $\bm{c}$. Again, a more general statements which covers
     the cases where there an odd number of unit cells in the
     periodic directions relates the topological invariant to
     the product of the net parity changes in the presence
     and absence of a $\pi$ flux through each of the holes of
     the system with periodic boundary conditions.

     Adiabatic cycles of insulators in 3d in class A (without
     TRS or any sub-lattice symmetry) are classified by a $Z$
     invariant. We consider a particular non-trivial adiabatic
     cycle in a system which is periodic in the $x, y$
     directions and semi-infinite in the $z$-direction, which
     has the feature that at $t=\pi$, the Hamiltonian
     corresponds to that of a strong topological
     insulator~\cite{roy_unpublished}. For $ \tilde{t} = t- \pi
     $ close to $0$, a low-energy effective continuum theory
     can be obtained by a Taylor expansion of the Hamiltonian
     around the point at which the gap closes, which in this
     case happens at $\bm{k}=(0,0,0)$ and $\tilde{t}=0$. One
     can then analytically solve for the boundary state
     spectrum and obtain an boundary Hamiltonian, as outlined
     for the 1d case. On doing this, we obtain a 2D Dirac
     boundary Hamiltonian of the form: \begin{equation}
     H_{\textrm{b}}(\tilde{t}) = \int dk {\begin{pmatrix}
     \psi_{1}\\ \psi_{2} \end{pmatrix}}^T \begin{pmatrix}
     \tilde{t}& -k_x+ i k_y \\ -k_x - i k_y & -
     \tilde{t} \end{pmatrix} \begin{pmatrix} \psi_{1}\\
     \psi_{2} \end{pmatrix} \nonumber \end{equation} Now let us
     consider the same adiabatic cycle, but in the presence of
     a constant magnetic field, $\mathbf{B}= (0,0,-B)$ where
     $B>0$. It is well known that for a 2d Dirac Hamiltonian,
     in the presence of a magnetic field, relativistic Landau
     levels are formed~\cite{Schakel_91_1428}.  The lowest
     Landau level has an energy, $ \lambda_{0}= \tilde{t}$ with
     a degeneracy of ${eB/h} $ states per unit area.  We see
     that, as $\tilde{t}$ increases, this lowest Landau level
     crosses zero from below at $\tilde{t}=0$.

    For a finite system, we may associate a Hall conductance
   with the surface state whose energy crosses zero by
   computing the Chern number in the space of eigenvalues of
   the magnetic translation operators. Since the states that
   cross the Fermi energy (which is set at 0) have an
   associated Hall conductance, we regard non-trivial
   topological adiabatic cycles of 3d insulators as defining
   adiabatic pumps of Hall conductance. If we consider a cycle,
   where the system at the start of the cycle is a strong 3d
   topological insulator with a surface which has an boundary
   gap induced by a TRS breaking surface perturbation, then the
   surface Hall conductance at the end of a cycle in the
   presence of a magnetic field, while remaining a half
   integer, changes by an integer amount.

  Non-trivial adiabatic cycles of superconductors in class C
  are classified by a $Z$ invariant. These superconductors have
  an $SU(2)$ gauge symmetry associated with spin-rotational
  symmetry.  Given a lattice Hamiltonian in this class of the
  form: $ H = \sum_{s_3,i,j} t_{i,j,s_3} a^{\dagger}_{i,s_3}
  a_{j,s_3} + \left( \sum_{i,j} \Delta_{ij}
  a^{\dagger}_{i,\uparrow}a^{\dagger}_{j,\downarrow} +
  \textrm{h.c} \right)$ where orbital indices have been
  suppressed, $i,j$ are site indices and $s_3 = 1, s_3 =-1$
  correspond to up-spins and down-spins respectively, we
  consider an associated Hamiltonian obtained by a generalized
  Peierls substitution $t_{ij,s_3} \rightarrow t_{ij,s_3} e^{-i
  s_3 \frac{e}{\hbar} (\int_i ^j \bm{A.dr} )} $, $\Delta_{ij}
  \rightarrow \Delta_{ij} e^{-i  \frac{e}{\hbar} (\int_i ^j
  \bm{A.dr} )}$. Here, $\bm{A}$ is the vector potential which
  corresponds to an uniform and commensurate magnetic field,
  $\mathbf{B}$ in the z-direction. The resulting Hamiltonian
  can be considered to be that of the system in a fictitious
  $SU(2)$ gauge field. Even in the presence of the $SU(2)$
  gauge, the Hamiltonian has a residual $U(1)$ symmetry for
  spin rotations about the $z$-axis.  The states that cross the
  zero of energy at a point where the gap at the boundary vanishes
  can thus be assigned a Hall conductance for spin along the
  z-direction. These adiabatic cycles can be
  therefore be interpreted as spin Hall pumps in the presence
  of an $SU(2)$ spin field.

   Adiabatic cycles of 0d systems can not be regarded as
   pumps. However, we note that the Berry phase associated with
   adiabatic cycles in classes AI and BDI can only be $0$ or
   $\pi$ and the latter correspond to the non-trivial adiabatic
   cycles. Similarly, the adiabatic cycles of insulators in
   classes AIII and C with an odd integer invariant also have a
   Berry phase of $\pi$.  This concludes our survey of some of
   the interesting examples of non-trivial adiabatic
   cycles. The current discussion did not cover adiabatic
   cycles in classes AIII, CII and CI, and those of 3d
   insulators in class AII which will be discussed
   elsewhere~\cite{roy_unpublished}.

  To summarize, we have studied the classification of adiabatic
  cycles of Bloch insulators. We showed that there exist
  non-trivial adiabatic cycles of trivial insulators, i.e.,
  paths in the spaces of Bloch Hamiltonians where the boundary
  gap must necessarily close though the bulk gap remains open
  and the insulators are themselves in the trivial class. We
  found that many of these adiabatic cycles can be regarded as
  pumps and that the net change in certain quantum numbers
  associated with the states that cross the gap equals the
  topological invariant describing the cycles.

The author is grateful to  Sudip Chakravarty, John Chalker,
Graeme Segal, Steve Simon, Shivaji Sondhi, Ady Stern and
Michael Stone for helpful discussions and gratefully
acknowledges support from EPSRC Grant No.  EP/D050952/1.
\bibliographystyle{h-physrev4}

\end{document}